\documentstyle[epsfig]{mn}
\def\beq{\begin{equation}}
\def\enq{\end{equation}}
\def\beqn{\begin{eqnarray}}
\def\enqn{\end{eqnarray}}

\begin{document}

\title{Possible TeV Source Candidates Among The Unidentified EGRET Sources}

\author[Wang, W. et al.]
{W. Wang$^*$, Z.J. Jiang, C.S.J. Pun, \& K.S. Cheng \\
Department of Physics, The University of Hong Kong, Pokfulam Road,
Hong Kong, China \\
$*$ present address: Max-Planck-Institut f\"ur
extraterrestrische Physik, Postfach 1312, 85741 Garching, Germany}

\maketitle

\begin{abstract}
We study the $\gamma$-ray emission from the pulsar magnetosphere
based on outer gap models, and the TeV radiation from pulsar wind
nebulae (PWNe) through inverse Compton scattering using a one-zone
model. We showed previously that GeV radiation from the
magnetosphere of mature pulsars with ages of $\sim 10^5-10^6$
years can contribute to the high latitude unidentified EGRET
sources. We carry out Monte Carlo simulations of $\gamma$-ray
pulsars in the Galaxy and the Gould Belt, assuming values for the
pulsar birth rate, initial position, proper motion velocity,
period, and magnetic field distribution and evolution based on
observational statistics. We select from the simulation a sample
of mature pulsars in the Galactic plane ($|b|\leq 5^\circ$) and a
sample at high latitudes ($|b|> 5^\circ$) which could be detected
by EGRET. The TeV fluxes from the pulsar wind nebulae of our
simulated sample produced through inverse Compton scattering by
relativistic electrons on the cosmic microwave background and
synchrotron seed photons are calculated. The predicted fluxes are
consistent with the present observational constraints. We suggest
that strong EGRET sources may be potential TeV source candidates
for present and future ground-based TeV telescopes.

\end{abstract}

\begin{keywords}
radiation mechanisms: nonthermal -- stars: statistics -- stars:
neutron -- pulsars: general -- $\gamma$-rays
\end{keywords}

\pagebreak

\section{Introduction}
There are about 170 unidentified $\gamma$-ray sources in the third
EGRET catalog, and nearly one third of these sources lie close to
the Galactic plane $|b|<5^\circ$ (Hartman et al. 1999). Most of
those unidentified sources in the Galactic plane can be identified
as $\gamma$-ray pulsars, possibly Geminga-like pulsars which are
radio quiet (Cheng \& Zhang 1998; Zhang, Zhang \& Cheng 2000). For
the medium and high latitude sources, it has been suggested that
some of them are associated with the supernova remnants in the
nearby Gould Belt (Gehrels et al. 2000; Grenier 2000). In
addition, Harding \& Zhang (2001) used the polar cap model
(Daugherty \& Harding 1996) to investigate if $\gamma$-ray pulsars
viewed at a large angle to the neutron star magnetic pole could
contribute to unidentified EGRET sources in the medium latitudes
associated with the Gould Belt. Their results suggest that at
least some of radio-quiet Gould Belt sources detected by EGRET
could be such off-beam $\gamma$-ray pulsars.

At the same time, these $\gamma$-ray pulsars could produce wind
nebulae through the interactions between relativistic wind
particles with the interstellar medium (ISM). The pulsar wind
nebulae will contribute to the production of non-pulsed X-ray
emission by synchrotron processes (Chevalier 2000), and TeV
photons through inverse Compton scattering (ICS)(Aharonian, Atoyan
\& Kifune 1997). These excess TeV photons have been detected in
some known pulsar wind nebulae, such as the Crab, Vela, PSR 1706-44,
possibly Geminga (Kifune et al. 1995; Yoshikoshi et al. 1997;
Aharonian et al. 1999; Lessard et al. 2000). Therefore, if
$\gamma$-ray pulsars contribute to the unidentified EGRET sources,
possible TeV signals could be expected to be detected in
these EGRET sources. Several groups have searched for TeV signals
in the error boxes of unidentified EGRET sources, for example,
with the HEGRA AIROBICC array (Aharonian et al. 2002a), and the
Whipple 10m Gamma-Ray Telescope. No TeV source detection has been
confirmed at Whipple, with only an upper limit TeV flux for
about 20 EGRET sources determined at $\sim (3-6)\times 10^{-11}\
{\rm photon\ cm^{-2}\ s^{-1}}$ (Fegan \& Weekes 2004). Deep
observations of the Cygnus region using HEGRA (Aharonian et al.
2002b) showed an unidentified TeV source in the vicinity of
Cygnus OB2 with an integral flux $F(>1\ {\rm TeV})\sim 5\times
10^{-13}\ {\rm photon\ cm^{-2}\ s^{-1}}$, at the edge of the 95\%
error circle of the EGRET source 3EG J2033+4188. However, no
connection between the EGRET source and the TeV signal has yet been
confirmed.

It has been shown that GeV photons can be produced in the pulsar
magnetosphere in outer gap models (Cheng et al. 1986; Zhang \&
Cheng 1997). A revised outer gap model (Zhang et al. 2004) takes
into account the effect of the inclination angle $\alpha$ between
the magnetic axis and the rotational axis, which can determine the
gap size of the outer gap. This allows some pulsars with
appropriate combinations of $\alpha, P$ and $B$, to maintain the
outer gap for at least $\sim~10^6$ years. Their advanced ages
allow these pulsars enough time to move up to high Galactic
latitudes as weak $\gamma$-ray sources. This leads Cheng et al.
(2004a) to propose that mature $\gamma$-ray pulsars with ages
$\sim 10^5-10^6$ years can contribute to the unidentified
EGRET sources. These mature pulsars also remain active in producing
relativistic wind particles, and form compact wind nebulae. In
addition, TeV photons can be created in the nebulae through the
ICS process. In this paper, we study the possible connection
between TeV $\gamma$-ray sources and the Unidentified EGRET
sources. We do not know if the Unidentified EGRET sources
are pulsars; even if they are pulsars we still do not know
their properties, i.e. period, magnetic field, inclination angle,
distance etc. Without these parameters we cannot calculate their
$\gamma$-ray properties. Therefore, we apply a statistical
approach using Monte Carlo simulation to study the Unidentified
Gamma-ray EGRET Sources. First we will simulate the galactic
pulsar population and use the outer gap model to calculate the
MeV-GeV photon power from these simulated pulsars. We have ignored
the contribution from the polar gap for simplicity. We can
determine which simulated pulsars can be detected by EGRET in
$\gamma$-rays; we call them $\gamma$-ray loud pulsars. The next
step is to calculate the $\gamma$-ray emission from the pulsar wind
based on the simulated pulsar parameters. We should point out that
the distribution of $\gamma$-ray loud pulsars is model dependent.
Subsequently we study the TeV $\gamma$-rays emitted from the pulsar
wind when they interact with their ambient interstellar medium. We
argue that strong EGRET sources may be potential TeV source
candidates for current and future TeV telescopes.

In \S 2, theories of radiation of mature pulsars from both inside
and outside the light cylinder will be briefly reviewed. A detailed
discussions of the ICS processes of relativistic electrons on the
background and synchrotron photons using a one-zone wind nebula
model (Chevalier 2000; Cheng, Taam \& Wang 2004b) will be
presented in \S~3. In \S~4, the properties of the identified TeV
pulsar wind nebulae sources are reviewed and compared with our
models. In \S~5, we describe the Monte Carlo simulations, similar
to those carried in Cheng et al. (2004a), for deriving the
distribution of $\gamma$-ray pulsars in the Galaxy and Gould Belt
which can be detected by EGRET. We then calculate the expected GeV
and TeV fluxes from these $\gamma$-ray pulsars generated by the
Monte Carlo simulations. The TeV flux distribution of the
unidentified EGRET source candidates, which could possibly be
detected by the present and future ground-based TeV telescopes,
will be presented in \S~6. Finally, we present our summary and
discussions in \S~7.

\section{Radiation theories associated with pulsars}
In this section, we will review the emission properties of mature
pulsars whose ages are from $\sim 10^5-10^6$ years.
The high energy radiation of mature pulsars can come both from the pulsar
magnetosphere inside light cylinder, and from the pulsar wind
nebula outside light cylinder.

\subsection{Inside light cylinder - outer gap model}
We present the $\gamma$-ray emission properties of pulsars using
the outer gap models originally proposed by Cheng et al.
(1986a,b). Based on the model, Zhang \& Cheng (1997) have
developed a self-consistent mechanism to describe the high energy
radiation from spin-powered pulsars. In their model, relativistic
charged particles from a thick outer magnetospheric accelerator
(outer gap) radiate through the synchro-curvature radiation
mechanism (Cheng \& Zhang 1996) rather than the synchrotron and
curvature mechanisms in general, producing non-thermal photons
from the primary $e^\pm$ pairs along the curved magnetic field
lines in the outer gap.

The characteristic emission energy of high energy photons emitted
from the outer gap is given by (Zhang \& Cheng 1997) \beq
E_{\gamma,c} \simeq 5\times 10^7 f^{3/2}B_{12}^{3/4}
P^{-7/4}({r\over R_L})^{-13/8}\ {\rm eV}, \enq where $P$ is the
rotation period, $B_{12}$ is the dipolar magnetic field in units
of $10^{12}$ G, $R_L=cP/2\pi$ is the light cylinder radius, $r$ is
the distance to the neutron star, and $f$ is the fractional size
of the outer gap. The $\gamma$-ray spectrum drops exponentially
beyond the energy $E_{\gamma,c}$.

The factor $f$, defined as the the ratio between the mean vertical
separation of the outer gap boundaries in the plane of the
rotation axis and the magnetic axis to the light cylinder radius,
is limited by the pair production between the soft thermal X-rays
from the neutron star surface and the high energy $\gamma$-ray
photons emitted from the outer gap region, and can be approximated
as $f\simeq 5.5 P^{26/21}B_{12}^{-4/7}$. The size of $f$ in turn
determines the total $\gamma$-ray luminosity of the pulsar, which
is given by (Zhang \& Cheng 1997) \beq L_\gamma \simeq f^3 L_{\rm
sd}, \enq where $L_{\rm sd}= 3.8\times 10^{31}B_{12}^2P^{-4}\ {\rm
erg\ s^{-1}}$ is the pulsar spin down power. However, the
estimation of gap size $f$ by Zhang \& Cheng (1997) does not
include the effect of inclination angle and their model can
produce very few pulsars with age $\sim 10^5-10^6$ yrs. Zhang
et al. (2004) have studied the properties of the outer gap by
including the effect of the inclination angle. Also, instead of taking
half of the light cylinder as the representation of the outer gap,
they have calculated the average outer gap size and use it as the
representation of the outer gap. This is very important for
a statistical study. Cheng et al (2004a) show that in Monte Carlo
simulations even if the inclination angle of pulsars is randomly
selected from a uniform distribution, the simulated $\gamma$-ray
pulsars detected by EGRET in the galactic plane are younger and
tend to have a larger inclination angle. On the other hand,
$\gamma$-ray pulsars detected by EGRET at higher latitudes are
older and have a smaller inclination angle. In other words, even
if the seed distribution has a uniform  distribution, a
non-uniform distribution can be generated by the $\gamma$-ray
selection effects. In this paper, we intend to use Monte Carlo
methods to study GeV and TeV $\gamma$-ray properties of pulsars, so
we adopt the model of Zhang et al. (2004) to determine the size of
the outer gap. Assuming that the representative region of the
outer gap is the average distance to the gap, the mean fractional
size of the outer gap can be approximated as $f(\alpha,P,B)
\approx \eta(\alpha,P,B)f(P,B)$, where $\eta(\alpha,P,B)$ is a
monotonically increasing function of $\alpha,\ B$, and $P$. The
value of $\eta$ roughly decreases by a factor 3 from large
inclination angles to smaller angles (Zhang et al. 2004). These
$\gamma$-rays from the pulsar magnetosphere will contribute to the
pulsed GeV photons in $\gamma$-ray pulsars detected by EGRET.

In general the differential energy spectrum of  $\gamma$-rays
for each pulsar is different. In principle we can calculate it if
the pulsar parameters are specified. However, it is very difficult
to do so in a Monte Carlo simulation, in which we need to deal with
over ten million pulsars. EGRET has detected six $\gamma$-ray
pulsars and their energy spectra from 100 MeV to a few GeV are all
very close to $E_\gamma^{-1}$ ( Hartman et al. 1999). For thin
outer gap pulsars, it has been shown that the energy spectrum is
proportional to $E_\gamma^{-1}log(E_{max}/E_\gamma) \propto
E_\gamma^{-1}$ (Cheng, Ho and Ruderman 1986b; Cheng and Ding
1994). For simplicity in the Monte Carlo simulation, we will
approximate the expected energy differential $\gamma$-ray flux of
the pulsar as \beq F(E_\gamma)\simeq {L_\gamma \over
\triangle\Omega d^2 E_\gamma}, \enq where $d$ is the distance of
the pulsar, and $\triangle\Omega$ is the the solid angle of
$\gamma$-ray beaming. The value of $\triangle\Omega$ generally
varies with different pulsars. For simplicity, we assume a
constant beaming solid angle $\triangle\Omega\sim 1$~sr in all our
analyses. In order to compare our model results with observations,
we calculate the expected integral EGRET flux of our model
$\gamma$-ray pulsars using the formula \beq
S_{\gamma}(E_{\gamma}\geq 100{\rm MeV})=\int_{\rm 100
MeV}^{E_{max}}F(E_\gamma)dE_\gamma , \enq where $E_{max}$ is the
maximum $\gamma$-ray energy detected by EGRET, chosen to be
50~GeV.

Here, we have ignored the $\gamma$-ray contribution from the polar
cap, which could be important for Unidentified EGRET Sources
(Gonthier et al. 2002). However, the phase-resolved EGRET data of
the Crab pulsar (Cheng , Ruderman and Zhang 2000), Geminga (Zhang
and Cheng 2001) and the Vela pulsar (Romani 1996) can be explained
very well by the outer gap model and it appears that polar cap
emission is unimportant. Most recently, Muslimov \&
Harding(2004) have suggested that the slot gap model could be able
to explain the phase-resolved data as well. Since the
distributions of simulated $\gamma$-ray pulsars are model
dependent, in this paper we will ignore the contribution of
$\gamma$-rays from the polar cap for simplicity.

\subsection{Outside light cylinder - pulsar wind nebulae}
Previous models of the pulsar wind nebulae
(e.g., Kennel \& Coroniti 1984; Chevalier 2000) mainly
concentrate on the bright nebulae produced by interactions between
young pulsar wind particles and the supernova remnant (SNR).
On the other hand, while there are no SNR surrounding the mature
pulsars, they remain active enough to produce the
relatively faint, compact synchrotron nebulae through interactions
between the relativistic wind particles and the interstellar medium.
In this work, we use the one-zone model to describe high
energy radiation from pulsar wind nebulae (Chevalier 2000).
In the model, the relativistic
electrons in the shock waves emit X-rays through synchrotron
radiation. This model can well explain the X-ray luminosity
and spectral properties of pulsar wind nebulae (Cheng et al. 2004b).

Mature pulsars move at a high proper velocity after their birth,
and can form a bow shock structure to produce synchrotron wind
nebulae when the pulsar proper motion velocity is larger than the
sound speed in the ambient interstellar medium (ISM). The
characteristic size of the shock wave produced by interactions
between the pulsar wind particles and the ISM is referred to as
the termination radius, $R_s$, and can be derived from \beq R_s =
(L_{\rm sd}/2\pi \rho v_p^2 c)^{1/2}\sim 10^{16}L_{\rm sd,
34}^{1/2}n^{-1/2}v_{p,350}^{-1}\ {\rm cm}, \enq where $\rho=nm_p$,
$n=1\ {\rm cm^{-3}}$ is the number density of the ISM, $m_p$ is the
proton rest mass, $v_p$ is the pulsar velocity in units of 350 km
s$^{-1}$ (cf. \S~5.2), and $L_{\rm sd, 34}$ is the spin down power
of the pulsar in units of $10^{34}$ erg s$^{-1}$.

We have assumed the pulsar wind can carry away most of the pulsar
spin-down power ($L_{sd}$) and deposit in the shock waves when the
wind interacts with the interstellar medium. In general, the
energy in the shock waves is stored in the magnetic field, the
energetic protons (ions) and electrons. However, the fractional
energy density of the magnetic field $\epsilon_B$ is low
(typically, $\epsilon_B\sim 0.001-0.01$, Kennel \& Coroniti 1984),
then by assuming equipartition of energy between protons and
electrons, we obtain $\epsilon_p \sim \epsilon_e\sim 0.5$. For a
given $\epsilon_B$, the magnetic field at the termination radius
is estimated as $B =(6\epsilon_B L_{\rm sd}/R_s^2 c)^{1/2}$. At
the shock front, the electron energy distribution is
$N(\gamma)\propto \gamma^{-p}$ for $\gamma_m <\gamma <\gamma_{\rm
max}$, where $\gamma_m = [(p-2)/(p-1)] \epsilon_e \gamma_w$, and
$\gamma_w$ is the Lorentz factor of the pulsar wind particles.

We derive the value of $\gamma_ {\rm max}$ by two methods.
First, an estimate for $\gamma_ {\rm max}$ can be obtained by equating the
synchrotron cooling timescale to the electron acceleration
timescale. The former timescale is given by $t_{\rm syn}=6\pi m_e
c/{\sigma_T\gamma B^2}$, and the latter timescale is given by
$t_{\rm acc}= \gamma m_ec/{eB}$, leading to $\gamma_{\rm
max,1}=(6\pi e/ {\sigma_TB})^{1/2}\sim 10^{10}B_{-5}^{-1/2}$,
where $\sigma_T$ is the Thompson cross section, and $B_{-5}$
denotes $B/10^{-5}{\rm G}$. Second, we can estimate $\gamma_ {\rm max}$
by equating the acceleration timescale and diffusion timescale of electrons. The
electron diffusion time in the termination radius can be estimated
as $t_{\rm diff}\sim (R_s/r_L)^2r_L/c$, where $r_L\sim \gamma
m_ec^2/eB$ is the Larmor radius.
This gives a maximum Lorentz factor of
$\gamma_{\rm max,2}\sim 10^8 B_{-5}R_{s,16}$,
where $R_{s,16}=R_s/10^{16}$ cm.
We will assume \beq \gamma_{\rm max}={\rm min}(\gamma_{\rm max,1},
\gamma_{\rm max,2}) \enq in the following discussion.

The Lorentz factor of wind particles cannot be too small, the
production of the synchrotron nebulae requiring, for example,
$\gamma_w>10^4$. Theoretical constraints on $\gamma_w$ are still
difficult to determine at present. In this paper, we use the
arguments of Ruderman (1981) and Arons (1983) that large fluxes of
protons (ions) can be extracted from neutron stars and can be
accelerated in the parallel electric field in the magnetosphere.
We also assume that the initial Poynting flux can be converted
into thermal and kinetic energy of the particles well within the
termination radius, and that the Lorentz factors of electrons and
protons are the same. For young pulsars like the Crab pulsar and
the Vela pulsar, a large number of electrons and positrons can be
produced in the outer magnetosphere (Cheng, Ruderman \& Ho 1986b)
and hence electrons/positrons are the main energy carriers in the
pulsar wind. However, for mature pulsars with age  $>10^5$ years,
the outer gap cannot produce too many pairs (Zhang \& Cheng
1997). The pair creation will be mainly at the polar cap, however, the
ratio between electron number and proton number is $\leq m_p/m_e$,
in particular for the space-charge-limited flow situation (Arons
\& Scharlemann 1979; Cheng \& Ruderman 1980). Since we will show
that the main contributors of TeV $\gamma$-ray sources are mature
pulsars, so we shall assume that protons are the main energy carriers
of the pulsar wind. We want to point out that this assumption does not
conflict with the equipartition assumption in the shock waves
because pulsar wind energy will be redistributed when the wind
interacts with the interstellar medium in the shock waves.

The protons can then carry away most of the spin down power in this
scenario (Coroniti 1990). Thus, we can derive the spin down power
as $L_{\rm sd}\sim \dot N \gamma_wm_pc^2$, where $\dot N$ is the
outflow current from the surface. This outflow current should be
of the same order as the Goldreich-Julian current (Goldreich \&
Julian 1969) which is given as $\dot N \simeq 1.35\times
10^{30}B_{12}P^{-2}\ {\rm s^{-1}}\propto L_{\rm sd}^{1/2}$. For
mature pulsars with $L_{\rm sd}\sim 10^{34-36}\ {\rm erg\
s^{-1}}$, we derive a typical value of $\gamma_w\sim 10^6$.

The critical synchrotron frequency of relativistic electrons with
a Lorentz factor $\gamma$ is $\nu(\gamma)=\gamma^2 eB/2\pi m_e c$.
The synchrotron radiation properties from the pulsar wind nebulae
depend on two characteristic frequencies: $\nu_m$, radiated by the
electrons with Lorentz factor  $\gamma_m$, and the cooling
frequency, $\nu_c$, given as $\nu_c= e/ (2\pi m_ecB^3)[(6\pi
m_ec)/(\sigma_T t_0)]^2$, where $t_0$ is the characteristic
timescale of the nebula estimated from the flow timescale in a
characteristic radiation region. If we define a characteristic
time scale $t_0\sim R_s/v_p\sim 10^{9}$~s, and $B\sim 10^{-5}$ G,
then the cooling frequency $\nu_c \sim 10^{19}$~Hz. This
corresponds to a characteristic cooling Lorentz factor
$\gamma_c\sim 10^{9}$, which is significantly larger than
$\gamma_m$. Therefore, we can derive the luminosity and spectral
properties of synchrotron radiation from the pulsar wind nebulae.
When $\nu_m<\nu<\nu_c$, it can be described as a non-thermal
spectrum with a power law photon index $\Gamma \sim (p+1)/2$, and
a synchrotron radiation luminosity estimated by (also see
Chevalier 2000; Cheng et al. 2004b) \beqn L_{\rm syn}(\nu)\simeq
{\sigma_Te^{(p-3)/2}(nm_p)^{(p+1)/4} \over 9
m_e^{(p-1)/2}c^{(p-1)/2}} \epsilon_e^{p-1} \epsilon_B^{(p+1)/4}
\nonumber \\
v_p^{(p+1)/2}\gamma_w^{p-2} t_0 L_{\rm sd}\nu^{(3-p)/2}. \enqn
This case is called the slow cooling regime. On the other hand in the
fast cooling regime when $\nu>\nu_c$, the emission spectrum will
steepen because of the cooling of relativistic electrons, the
photon index $\Gamma \sim (p+2)/2$, and the luminosity is \beqn
L_{\rm syn}(\nu)\simeq {1\over 2}({p-2\over p-1})^{p-1}({e\over
m_e c})^{(p-2)/2}(nm_p)^{(p-2)/4} \epsilon_e^{p-1} \nonumber \\
 \epsilon_B^{(p-2)/4} \gamma_w^{p-2} v_p^{(p-2)/2}L_{\rm
sd}\nu^{(2-p)/2}. \enqn

\section{TeV photons from pulsar wind nebulae through inverse Compton processes}
The synchrotron radiation of relativistic electrons in the pulsar
wind nebulae can well contribute to non-thermal radio to X-ray
emissions. At the same time, the inverse Compton scattering (ICS)
of the same relativistic electrons on the ambient photon fields
results in the production of the very high energy TeV photons
observed (de Jager \& Harding 1992; Atoyan \&~Aharonian 1996;
Aharonian, Atoyan \&~Kifune 1997). In this section, we will
discuss two cases of the ICS processes of two different types of
seed photons, namely, background and synchrotron photons. We can
see that the Inverse Compton luminosity is sensitively dependent
on the properties of the seed photons.

\subsection{Inverse Compton Scattering from background photons}

The relative magnitude between the emission from synchrotron and
ICS processes depends on the magnetic field $B$ and the photon
field density $w_{\rm ph}$. The magnetic field in the interstellar
medium is typically within the range $(0.3-1)\times 10^{-5}$~G.
Synchrotron X-ray photons and $\gamma$-ray photons from ICS have
been shown to be produced in regions of different magnetic fields
(Aharonian et al. 1997). In this paper, we assume for simplicity
that the X-rays and $\gamma$-rays come from the same region with a
magnetic field $B\sim 10^{-5}$~G.

The photon energy density surrounding the pulsars can be
attributed to several origins: first, the cosmic
microwave background (CMB) radiation contributes a constant energy
density $w_{\rm CMB}\sim 4\times 10^{-13}{\rm erg\ cm^{-3}}$, second,
the diffuse galactic dust far-infrared (FIR) and star light
near-infrared and optical background (sl), and third, possible radiation
fields of local origin. The density of the galactic background
field varies from site to site, with an average values
$w_{\rm FIR}\sim 10^{-13}\ {\rm erg\ cm^{-3}}$ and $w_{\rm sl}\sim
10^{-12}\ {\rm erg\ cm^{-3}}$ (Mathis, Metzger \& Panagia 1983).
Although the energy density of the galactic radiation background
is higher than that of the CMB, its importance is strongly reduced in
the energy region above 100~GeV due to the Klein-Nishina effect
(Aharonian et al. 1997). Therefore we consider here only the
CMB photon field as the external seed photons for the
production of TeV photons.

The characteristic energy, $E_{\rm syn}$, of synchrotron radiation
from relativistic electrons with typical Lorentz factor $\gamma$
is given as \beq E_{\rm syn}=h\nu_{\rm syn}=\hbar \gamma^2eB/m_ec
\sim 1B_{-5}(\gamma/10^8)^2 \ {\rm keV}. \enq The characteristic
energy of the photons from the ICS of these same energy electrons
on the CMB photons is given by \beq E_{\rm IC}=\gamma^2 h\nu_0\sim
1(\gamma/10^8)^2(h\nu_0/\varepsilon_{\rm CMB})\ {\rm TeV}, \enq
where $\varepsilon_{\rm CMB}\sim 10^{-4}$~eV is the energy of the
CMB photons. The ratio of the luminosity of the emission from the
ICS process to that from synchrotron radiation can be simply
obtained as ($\gamma h\nu_0<m_e c^2$) \beq {L_{\rm IC}\over L_{\rm
syn}} = {w_{\rm CMB}\over w_B}, \enq where $w_B=B^2/8\pi$ is the
energy density of the magnetic field. To derive the $\gamma$-ray
luminosity of TeV photons by the ICS process, we can first
estimate the synchrotron X-ray luminosity of the pulsar wind
nebulae using the one-zone model discussed in \S~2.2.

The TeV photon luminosity and flux distributions will be studied
in detail in \S~6 using a simulated sample of mature pulsars.
Here, we use some typical pulsar parameters for a simple estimate.
According to equations (9) and (10), TeV
photons through ICS produced by relativistic electrons correspond
to X-rays ($\sim 1$ keV) through synchrotron processes. In the
calculation, $\nu\sim 10^{18}$ Hz denotes the X-rays through the
synchrotron process, and we take some typical parameters of
pulsars and nebulae to find the TeV luminosity: $L_{\rm sd}\sim
10^{34}\ {\rm erg\ s^{-1}}, \epsilon_e\sim 0.5, \epsilon_B\sim 0.01,
\gamma_w\sim 10^6, R_s\sim 10^{16}$ cm, the electron energy
spectral index $p=2.2$, then we obtain the TeV luminosity by ICS
processes $L_{\rm IC}({\rm TeV})\sim 10^{31}\ {\rm erg\ s^{-1}}$.

The spectral profile is also related to the synchrotron process.
As discussed in \S 2.2, there exist three Lorentz factors of the
relativistic electrons $\gamma_m, \gamma_{max},$ and $\gamma_c$.
The Lorentz factor of the ICS electrons should satisfy $\gamma
h\nu_{\rm CMB}<m_e c^2$, which gives a critical Lorentz factor
$\gamma_{cut}\sim 10^9$. Then when the Lorentz factors satisfy the
relation $\gamma_m<\gamma_c<\gamma_{cut}\leq\gamma_{max}$, for ICS
electrons with $\gamma$ between $\gamma_m$ and $\gamma_c$, the ICS
spectrum will show a power-law with a photon index $\Gamma =
(p+1)/2$, for $\gamma$ between $\gamma_c$ and $\gamma_{cut}$, the
photon index becomes $\Gamma=(p+2)/2$. For ICS electron $\gamma$
above $\gamma_{cut}$ or $\gamma_{max}$, the spectrum will become
an exponential decay form reflecting a high energy cut-off. For
mature pulsars discussed above with $\gamma_{max}\sim 10^8$, a
cut-off energy at $\sim$1~TeV will be produced.

\subsection{Synchrotron self-Compton scattering contribution}
Using the synchrotron spectrum as the sources of seed photons, we
can compute the resulting ICS emission (called synchrotron
self-Compton, SSC, photons). Similar to the original synchrotron
spectrum, the up-scattered component also has two
characteristic energies, $E_m^{\rm SSC}\sim h\gamma_m^2\nu_m$, and
$E_c^{\rm SSC}\sim h\gamma_c^2\nu_c$. In addition, the condition
$\gamma h\nu(\gamma)< m_ec^2$ will give a characteristic critical
factor $\gamma_{\rm crit}\sim 2\times 10^6$ for $B=10^{-5}$~G and
a critical energy $E_{\rm crit}\sim \gamma_{\rm crit}m_ec^2$. Due
to the limitation that $\gamma_{max} h\nu_0<m_e c^2$, we can find
a frequency $\nu_0$ such that if $\nu_0>\nu_m$, $E^{\rm
SSC}_{cut}\sim \gamma_{max}m_ec^2\ \sim
50(\gamma_{max}/10^8)$~TeV. In the case of $\nu_0<\nu_m$, a lower
limit of $\gamma'_{m}h\nu_m\sim m_ec^2$ is required, thus implying
a cut-off energy $E^{\rm SSC}_{cut}\sim \gamma'_{m}m_ec^2\sim
10$~TeV for $B\sim 10^{-5}$~G and $\gamma_m\sim 10^5$. If
$E_m^{\rm SSC}<E_c^{\rm SSC}<E^{\rm SSC}_{cut}$, then the SSC
spectrum will show a power-law with a photon index $\Gamma =
(p+1)/2$ for photon energy below $E_c^{\rm SSC}$, and an index
$\Gamma=(p+2)/2$ for photon energy above $E_c^{\rm SSC}$. For
photon energy $> E\sim \gamma_{crit} m_ec^2$, the energy spectrum
of SSC becomes steeper because of the Klein-Nishina effect
(Aharonian et al. 1997).

In general, the intensity of the ICS emission relative to the
synchrotron radiation can be estimated by considering the ratio of
the energy density of synchrotron radiation and the magnetic
field, $w_{\rm syn}/w_B$. In our one-zone model, synchrotron
radiation comes from the termination radius $R_s$. We approximate
the synchrotron energy density as $w_{\rm syn}\sim 3L_{\rm
syn}t_f/4\pi R_s^3$, where $t_f\sim R_s/c$. The SSC photon energy
is given by $E_{\rm SSC}\sim \gamma^2 h\nu$, where $\nu\sim
\gamma^2 eB/m_e c$. Because the relativistic electrons producing
the TeV photons by SSC correspond to a characteristic Lorentz
factor $\gamma\sim 10^6 <\gamma_c$ (assuming $B\sim 10^{-5}$ G and
$t_0\sim 10^9$~s), we should apply equation (7) to compute the
energy density and luminosity of low energy synchrotron seed
photons. The ratio of the SSC and synchrotron luminosities can be
given by \beqn {L_{\rm SSC}\over L_{\rm syn}}\simeq
{\sigma_Te^{(p-3)/2}(nm_p)^{(p+1)/4} \over 9
m_e^{(p-1)/2}c^{(p-1)/2}} \epsilon_e^{p-1} \epsilon_B^{(p-3)/4}
\nonumber \\ v_p^{(p+1)/2}\gamma_w^{p-2} t_0 \nu_{\rm
seed}^{(3-p)/2}. \enqn Assuming $\nu_{\rm seed}\sim 10^{14}$~Hz,
$t_0\sim 10^{9}$~s, and taking the other parameters the same as
those in \S~3.1, we determine the luminosity of TeV photons
produced by SSC as about $L_{\rm SSC}({\rm TeV})\sim 10^{30}\ {\rm
erg\ s^{-1}}$.

\section{Identified TeV sources of pulsar wind nebulae}
The present TeV telescopes have detected excess TeV photons
from three $\gamma$-ray pulsars, namely, Crab, Vela and PSR
1706-44, with a possible detection from Geminga (see reviews by
Aharonian 1999; Weekes 2004). Their $\gamma$-ray emissions in the
GeV band are dominated by contributions from the pulsar
magnetosphere, as predicted in the outer gap model (\S~2.1).
However, no pulsed TeV photons are detected by the present
telescopes (Kifune et al. 1995; Yoshikoshi et al. 1997; Aharonian
et al. 1999; Lessard et al. 2000), thus implying that the TeV
photons mainly come from their wind nebulae. In this section, we
check our one-zone model by estimating the TeV signals from these
four sources. Because of the simplified nature of these models, we
do not expect to reproduce the detailed properties of the nebula.
More detailed discussions for individual sources have been
addressed by other authors (e.g. de Jager \& Harding 1992; Atoyan
\&~Aharonian 1996; Aharonian, Atoyan \&~Kifune 1997).

\subsection{Crab nebula}
The Crab nebula is the first pulsar wind nebula with TeV signal detection
(Weekes et al. 1989), and it has been extensively studied by many
TeV detectors since, for example, Milagro (Atkins et al. 2003),
HEGRA (Aharonian et al. 2000), Whipple (Lessard et al. 2000),
Tibet (Amenomori et al. 1999), and CANGAROO (Tamimori et al. 1998).
Detailed spectral and flux information have been obtained.
The Crab TeV spectrum can be fitted in the energy range above 1~TeV
by a simple power law with a photon index $\sim~2.5\pm 0.2$,
and shows no cut-off up to 100~TeV. The TeV luminosity
is estimated to be $\sim~10^{34}\ {\rm erg\ s^{-1}}$.

Atoyan \& Aharonian (1996) discussed the radiation mechanism of the
Crab nebula based on the model of Kennel \& Coroniti (1984), and
fitted the observed data from radio to GeV by synchrotron
radiation, and the TeV band through ICS processes.
With a complicated spatial structure, the Crab nebula probably
cannot be fully described in our one-zone model. What we present
in the following should therefore be treated as a rough check.

The magnetic field in the Crab nebula is relatively high at
$B\sim 10^{-3}-10^{-4}$~G (Hester et al. 1996), which significantly
reduces the relative contributions from the ICS processes.
With a ratio $w_{\rm CMB}/w_B\leq 10^{-4}$, the SSC contribution to
TeV photons should be the dominant component.
Assuming $B\sim 10^{-4}$~G, we derive the cooling
Lorentz factor of electrons to be $\gamma_c\sim 10^6$, and the
characteristic timescale $t_0\sim 3\times 10^{10}$~s.
The implied cooling energy $E_c^{\rm SSC}\sim 1$~TeV, which is
only 1\% of the predicted SSC cut-off.
Therefore, above 1~TeV, the photon index of the ICS spectrum is given by
$\Gamma=(p+2)/2$, where $p>2$.
On the other hand, the observed TeV spectrum of the Crab nebula implies
that $p\sim 3$. However, from the
observed synchrotron X-ray spectrum of the Crab nebula (Willingale
et al. 2001), $\Gamma=(p+2)/2\sim 2.1$, which gives $p\sim 2.2$.
Since $\gamma_{crit}\sim \gamma_c\sim 10^6$ in the Crab nebula,
and above $\gamma_{crit} m_ec^2\sim 1$~TeV, the SSC spectrum can be
steeper than $(p+2)/2$, hence agreeing with the observations.
The discrepancy may also result from the spatial variations of the
electron spectrum and magnetic field in the Crab nebula considered
by other models (de Jager \&~Harding 1992; Atoyan \&~Aharonian 1996;
Sollerman et al. 2000), which are not considered in our one-zone model.

Despite the very large size of the Crab nebula, the TeV photons
come from a central region $\sim 1$~pc (Atkins et al. 2003). The
termination radius of the nebula is about $R_s=4\times
10^{17}$~cm. In our one-zone model, we can take
$\epsilon_e\sim\epsilon_B\sim 0.5, \ p=2.2$, and $\gamma_w\sim
10^7$, to estimate the synchrotron luminosity and energy density.
At $\nu\sim 10^{14}$ Hz, the luminosity $L_{\rm syn}\sim 10^{36}\
{\rm erg\ s^{-1}}$ and $w_{\rm syn}/w_B\sim 0.03$. Therefore the
SSC contribution to TeV luminosity is $L_{\rm SSC}\sim 3\times
10^{34}\ {\rm erg\ s^{-1}}$, which is comparable to the observed
value.

\subsection{Vela and PSR 1706-44}
Yoshikoshi et al. (1997) used the 3.8 m imaging Cerenkov telescope
near Woomera, South Australia to detect TeV photons of the Vela
pulsar region. The detected TeV $\gamma$-ray flux is $2.9\times
10^{-12}{\rm photon\ cm^{-2}\ s^{-1}}$, corresponding to a
luminosity of 3$\times 10^{32}\ {\rm erg\ s^{-1}}$ assuming the
distance of Vela to be 300 pc. Similar to Vela, PSR 1706-44 has
also been detected in the TeV energy range (Kifune et al. 1995;
Chadwick et al. 1998). The flux above a threshold of 1 TeV is
$\sim 10^{-11}\ {\rm photon\ cm^{-2}\ s^{-1}}$. However, recent
observations of HESS reported non-detection of PSR 1706-44, with
an upper limit flux $\sim 1.3\times 10^{-12}{\rm erg\ cm^{-2}\
s^{-1}}$ (Aharonian et al. 2005). The reason for the observational
discrepancy is unknown, possibly the TeV emission could be
variable on a time scale of years, e.g. in the different medium
environments. Assuming a distance of 1.5 kpc, the TeV luminosity
is lower than $4\times 10^{32}\ {\rm erg\ s^{-1}}$. TeV spectral
information has been obtained for neither of these two sources.

Vela and PSR 1706-44 have nearly the same spin down power of a few
times $10^{36}\ {\rm erg\ s^{-1}}$. This suggests that their
nebulae have similar properties moving in the interstellar medium.
Assuming $\epsilon_e \sim 0.5$ and $\epsilon_B \sim 0.001$, we
estimate the magnetic field $B$ in the two pulsar wind nebulae.
According to the Chandra observation of Vela (Pavlov et al. 2001),
together with $R_s\sim 10^{17}$~cm (Cheng et al. 2004b), and
$L_{\rm sd}\sim 6.9\times 10^{36}\ {\rm erg\ s^{-1}}$, we estimate
that $B\sim 10^{-5}$~G. For PSR 1706-44, using $L_{\rm sd}\sim
3.4\times 10^{36}\ {\rm erg\ s^{-1}}$, and $R_s\sim 3\times
10^{17}$~cm (Finley et al. 1998; Cheng et al. 2004b), we estimate
$B\sim 3\times 10^{-6}$~G. The observed nebula photon indices of
Vela and PSR 1706-44 in X-rays is about 1.7, which requires
$\nu_c>\nu_x$, and $p\sim 2.4$. As discussed in \S~3.2, the ICS
processes by CMB photons would be the dominant contributor to the
TeV flux of the two pulsar wind nebulae. The proper motion
velocity of Vela is about 65 km~s$^{-1}$ (Pavlov et al. 2001).
Assuming for the two pulsars $t_0\sim 10^{10}$~s, $\gamma_w\sim
10^6$, and $\nu\sim 3\times 10^{17}$~Hz, to find the synchrotron
radiation, we estimate the TeV luminosities of the pulsar nebulae
to be $\sim 5\times 10^{32}\ {\rm erg\ s^{-1}}$ for Vela, and
$\sim 2\times 10^{33}\ {\rm erg\ s^{-1}}$ for PSR 1706-44. So the
present assumption of model parameters for two pulsar wind nebulae
may predict higher luminosities than are observed.

\subsection{Geminga}
Geminga is a strong $\gamma$-ray pulsar source in the GeV energy
range, but is very weak in other energy bands. For example, its
X-ray luminosity (including pulsar and the wind nebula) is about
$\sim 10^{30}\ {\rm erg\ s^{-1}}$ (Caraveo et al. 2003).
Observations of Geminga have not been yielded a significant excess
of TeV photons. Aharonian et al. (1999) obtained an upper limit
which is only $\sim$~13\% of that of the Crab nebula,
corresponding to a luminosity of $<6\times 10^{30}\ {\rm erg\
s^{-1}}$ assuming a distance of 160 pc.

The wind nebula around Geminga has a typical bow shock structure,
with a non-thermal X-ray spectrum and a photon index $\Gamma\sim
1.6$, which corresponds to $p\sim 2.2$ (Caraveo et al. 2003).
Assuming that the Geminga pulsar moves in the interstellar medium
of density $n\sim 1\ {\rm cm^{-3}}$ with a proper motion velocity
of $v_p\sim$120\ km\ s$^{-1}$ (Bignami \& Caraveo 1993), and a
spin down power of $L_{\rm sd} \simeq 3.2\times 10^{34}\ {\rm erg\
s^{-1}}$, the termination shock radius of the pulsar wind nebula
can be determined as $R_s \sim 4\times 10^{16}$~cm, with $t_0\sim
3\times 10^6$~s. Taking $\gamma_w\sim 10^6$, $\epsilon_e \sim
0.5$, $\epsilon_B \sim 0.01$, and $B\sim 10^{-5}$~G, Cheng et al.
(2004b) explained the X-ray features of Geminga with a one-zone
model. The SSC contribution to the TeV flux of Geminga is also very
small. The total ICS luminosity of the nebula in the TeV energy
range is estimated as $\sim 3\times 10^{29}\ {\rm erg\ s^{-1}}$,
which is lower than the upper limit inferred by observations.

\section{Simulation of $\gamma$-ray pulsars in the Galaxy}
In this section we briefly describe the Monte Carlo method, which
is used to simulate the properties of the luminosity, spatial
evolution, and distributions of $\gamma$-ray pulsars in the
Galaxy. The detailed Monte Carlo steps can be found in Cheng \&
Zhang 1998; Zhang et al. 2000; Fan et al. 2001).

\subsection{Monte Carlo simulation of $\gamma$-ray pulsars in the Galaxy}
The basic assumptions of our Monte Carlo simulation are given as
follows:

1. The birth rate of pulsars in the Galaxy is not clear, it is
about one pulsar every 50-200 years. Here we use the pulsar
birth rate in the Galaxy is one pulsar per century. The age of the
Gould Belt is $\sim 3\times 10^7$ yrs. Again the birth rate of pulsars in
the Gould Belt is not certain but a rate of 20 Myr$^{-1}$ is generally
used (Grenier 2000). Basically the birth rates will only affect
the final number of $\gamma$-ray pulsars but not their
distributions.

2. The Sun is inside the Gould Belt and is located about 200 pc
towards $l=130^\circ$ (Guillout et al. 1998). The Gould Belt has
an ellipsoidal shaped ring with semi-major and minor axes equal to
500 pc and 340 pc respectively.

3. The initial position for each pulsar in the Galaxy is
distributed according to the mass distribution of the Galaxy. We
use the mass distribution suggested by Paczynski (1990) and
Sturner \&~Dermer (1996). But the initial position of each pulsar
is assumed to be born uniformly inside the Gould Belt.

4. The initial magnetic field $\log B$ is assumed to be Gaussian,
with a mean value of 12.5 and a dispersion of 0.3. Since the
pulsars in our simulation sample are all younger than 10 million
years, and the magnetic field does not decay in 10 Myr
(Bhattacharya et al. 1992), we have ignored the field decay for
these pulsars.

5. The initial period is chosen to be $P_0=10$~ms, and the period
at time $t$ is given by $P(t)=(P_0+1.95\times
10^{-39}B^2t)^{1/2}$.

6. The initial velocity of each pulsar is the vector sum of the
circular rotation velocity at the birth location and the random
velocity from the supernova explosion (Paczynski 1990). The
circular velocity is determined by the Galactic gravitational
potential and the Maxwellian three-dimensional root-mean-square
(rms) velocity is assumed to be $\sqrt{3}\times 100\ {\rm km\
s^{-1}}$ (Lorimer et al. 1997). Furthermore, the pulsar position
at time $t$ is determined following its motion in the Galactic
gravitational potential. Using the equations given by Paczynski
(1990) for the given initial velocity, orbit integrations are
performed using the fourth-order Runge Kutta method with variable
time steps on the variables $R, V_R, z$, and $V_z$. The sky
positions and distances to the simulated pulsars can then be
calculated.

7. The inclination angle $\alpha$ of each pulsar is chosen
randomly from a uniform distribution (Biggs 1990).

8. The $\gamma$-ray background is non-uniform over the sky,
therefore the $\gamma$-ray threshold varies over the sky as well.
In general the threshold will be higher in the galactic plane and
lower at higher latitudes. It is generally accepted that a detectable
source should have a signal at least $5\sigma$ above the
background, which is roughly equivalent to the likelihood
criterion $\sqrt{TS}\geq 5$(Hartman et al. 1999). Recently,
Gonthier et al. (2002) have estimated that photon flux thresholds
for EGRET are $1.6 \times 10^{-7}{\rm cm^{-2}~s^{-1}}$ for sources
located at $|b|<10^\circ$ and $7 \times 10^{-8}{\rm
cm^{-2}~s^{-1}}$ for sources located at $|b|>10^\circ$
respectively. In our simulation, we use energy flux threshold
instead of photon flux threshold. For simplicity, we have assumed
photon index $\sim 2$ for all simulated sources. We convert the
photon flux thresholds suggested by Gonthier et al. (2002) to
energy flux threshold as $\sim 1.5 \times 10^{-10}{\rm
erg~cm^{-2}~s^{-1}}$ for sources located at $|b|<10^\circ$ and
$\sim 6.8 \times 10^{-11}{\rm erg~cm^{-2}~s^{-1}}$ for sources
located at $|b|>10^\circ$ respectively.

\subsection{Simulation results}
We carry out Monte Carlo simulation of the Galactic pulsars born
during the past 10~Myr. We find a total of 76 $\gamma$-ray mature
pulsars of ages larger than $10^5$ years that could be
detected by EGRET. Out of this simulated sample, 44 of them lie in
the Galactic plane ($|b|\leq 5^\circ$) and 32 lie at higher
latitudes ($|b|> 5^\circ$). This current result is slightly
different from the simulations presented in Cheng et al. (2004a)
because of the revised age restrictions here of taking only mature
pulsars of $\geq 10^5$ years. Currently, four radio pulsars with
age $>10^5$ yrs are identified as $\gamma$-ray pulsars, i.e.
Geminga, PSR B1055-52, PSR B1951+32, PSR J0218+42 (Thompson et al.
1996; Kuiper et al. 2000). The predicted $\gamma$-ray pulsar
numbers appear very much larger than the confirmed $\gamma$-ray
pulsars. We should notice that first, not all 76 predicted are
radio-loud. The radio beaming factor is roughly 0.15 (Emmering and
\& Chevalier 1989; Biggs 1990). Taking the radio beaming factor
into account, we predict that $\sim$12 EGRET Unidentified Sources
will be identified in the radio band in future. In fact recently there
are 20 known radio pulsars located within the error boxes of EGRET
sources and many of them may be identified as radio-loud
$\gamma$-ray pulsars (Manchester 2004). It is very important to
note that TeV $\gamma$-rays from pulsar wind nebulae are isotropic
and independent of radio-loud or radio-quiet $\gamma$-ray
pulsars.

In Figure 1, we plot the distributions of the pulsar period (upper
panel) and the magnetic field (bottom panel) of our simulated
sample, for both sources in the the high latitudes of $|b|>
5^\circ$ (solid histogram), and in the low latitudes $|b|\leq
5^\circ$ (dashed histogram). The proper motion velocity (top
panel) and distance (bottom panel) distributions of our mature
pulsar sample are presented in Figure 2. The average velocity of
our sample is found to be $\sim 350\ {\rm km\ s^{-1}}$. We have
taken this value in equation (5) to estimate the termination
radius (cf. \S~2.2). We show also in Figure~2 that the low and
high latitude pulsar samples have different distance
distributions. The average distance of the pulsars in the Galactic
plane ($|b|\leq 5^\circ$) is determined to be $<d>\simeq 900$~pc,
while the average distance is $<d>\simeq 400$ pc for the high
latitude sample ($|b|> 5^\circ$). This suggests that many high
latitude $\gamma$-ray mature pulsars may lie in the Gould Belt.

In Figure 3, we plot the $\gamma$-ray luminosity versus spin down
power of our simulated $\gamma$-ray pulsar sample. The spin down
power are all found to be lower than $10^{36}\ {\rm erg\ s^{-1}}$,
which is consistent with our expectations. Many of them in the
Galactic plane may be the Geminga-like pulsars. If we fit the
simulated relation of $\gamma$-ray luminosity versus spin down
power with single power law and it is consistent with the observed
correlation of the known $\gamma$-ray pulsars, $L_{\gamma, \rm
GeV}\propto L_{\rm sd}^{0.5}$ (Thompson 2001). However, from
Figure 3 there probably exist two populations of the relations:
one population follows $L_\gamma \propto L_{sd}$; another
population emerges as a branch for $L_{sd} > 10^{34}{\rm erg\
s^{-1}}$, with $L_\gamma \propto L_{sd}^{1/4}$. Zhang et al.
(2004) have provided a possible explanation for the nature of these two
populations and they estimate the separation of the two populations
should occur at $1.5\times 10^{34}{\rm P^{1/3} erg ~s^{-1}}$.

\section{TeV source candidates of the unidentified EGRET sources}
Assuming that mature pulsars can form faint and compact nebulae,
and can produce TeV photons through ICS processes, the ICS flux
from the pulsar nebulae can be calculated by \beq F_{\gamma, \rm
TeV}=L^{\rm ICS}/4\pi d^2 E_\gamma, \enq where $L^{\rm ICS}\simeq
L_{\rm IC}+L_{\rm SSC}$ (see \S~3), and $E_\gamma$ is the
threshold energy, taken to be 1~TeV. In order to calculate the TeV
luminosity for each of the simulated pulsars. we have to choose some
fixed values for the pulsar wind parameters, i.e. $\epsilon_e,
\epsilon_B$, p and $\gamma_w$ , which are produced in non-linear
processes and will certainly vary from pulsar to pulsar. We have
assumed a set of pulsar wind nebula parameters for all simulated
pulsars, with $\epsilon_e\sim 0.5$ according to the energy
equipartition assumption, $\epsilon_B\sim 0.01$, $p\sim 2.2$,
$\gamma_w\sim 10^6$, and the ISM density $n\sim 1$~cm$^{-3}$. The
choices of these parameters are consistent with theoretical
estimations.

In Figure 4, we plot the GeV $\gamma$-ray photon flux from the
pulsar outer gap versus the TeV photon flux from the pulsar wind
nebula. There exists a correlation between the GeV and TeV fluxes
because both quantities depend on the pulsar spin down power. This
suggests that strong EGRET sources may be potential TeV sources to
be detected by present and future TeV telescopes.

The TeV flux distribution of the mature pulsars at high latitudes,
$|b|> 5^\circ$ (solid histogram) and on the Galactic plane,
$|b|\leq 5^\circ$ (dashed histogram), are presented in Figure~5.
The predicted TeV fluxes of our low latitude simulated sample are
all lower than $3\times 10^{-12}\ {\rm photon\ cm^{-2}\ s^{-1}}$,
which is also the upper flux limit of all but two of the high
latitude sample. The predicted TeV flux from our sample is lower
than the previous observational constraints mentioned earlier in
the Introduction. However, with the rapid advancements of the
ground-based TeV telescopes, some of the unidentified EGRET
sources could be identified as TeV sources in the future.

\section{Summary and discussion}

We study the high energy radiation from mature pulsars with
ages of $\sim 10^5-10^6$ years in the present paper. We
consider 100~MeV to GeV $\gamma$-rays generated from the
magnetosphere of mature $\gamma$-ray pulsars based on a new
self-consistent outer gap model, which includes the effects of
inclination angle and the average properties of the outer gap (Zhang
et al. 2004). The relativistic wind particles from these mature
pulsars interact with the interstellar medium, and form compact
wind nebulae. The wind nebulae then produce X-ray emission through
synchrotron radiation, and TeV photons from ICS by relativistic
electrons on cosmic microwave background and synchrotron seed
photons. We investigate the nebula radiation using a one-zone
model, which has been shown to be able to predict the high energy
properties of known pulsar wind nebulae (Cheng et al. 2004b). We
conclude that in mature pulsars, the synchrotron radiation and the
ICS processes like SSC cannot significantly contribute in the 100
MeV to GeV energy band because of the weak magnetic field present
($B\sim 10^{-5}$~G). However, it is possible that the nebulae of
young pulsars in the galactic plane surrounded by supernova
remnants, e.g. the Crab nebula, can contribute to the 100 MeV to
GeV energy band through synchrotron and SSC processes.

We would like to remark that pulsed TeV emission is expected to be
produced inside the light cylinder via inverse Compton scattering
between the soft photons and the relativistic charged particles
inside the outer gap. In the new outer gap models (Zhang et al.
2004), the soft photon density is just enough to convert one out
of 10$^5$ gamma-ray photons into pairs. The inverse Compton
scattering efficiency is expected to be lower than 10$^{-5}$ of
the outer gap power. Therefore, it is not surprised that none of
known EGRET pulsars has been detected with pulsed TeV photons. On
the other hand, TeV photons associated pulsars should be produced
by PWN via the inverse Compton scattering and therefore should be
unpulsed. The Crab nebula has been  confirmed with TeV detection.
PSR1706-44 is considered as another confirmed TeV source (Weekes
2004) but the recent HESS results challenge this claim. Geminga is
a very nearby known EGRET pulsar but yet no positive detection on
TeV. For the former pulsar, we have speculated that the emission
from PWN may be variable with a time scale of years and the latter
may attribute to the fact that the magnetic field strength in PWN
is so weak that the TeV flux produced by SSC is also very weak. In
short, there are some evidences that TeV photons can be emitted
from some known EGRET pulsar PWNe. Therefore we believe that if
the EGRET unidentified sources are indeed pulsars, some of their
PWNe should emit detectable TeV photon fluxes.

Finally, we calculate the TeV fluxes from wind nebulae of these
mature pulsars through inverse Compton processes. The predicted
TeV fluxes are consistent with the recent observational
constraints obtained at the Whipple Observatory. In addition, our
results predict that the unidentified EGRET sources, especially
the strong sources, can be potential TeV sources to be be detected
by future ground-based TeV telescopes. Several third generation
ground-based TeV telescopes are under construction or recently
finished, e.g., MAGIC (Lorenz 1999; 2004), HESS (Hofmann 1999;
Hinton 2004), VERITAS (Weekes et al. 2002; Krennrich et al. 2004),
and CANGAROO-III (Matsubara 1997; Kubo et al. 2004). The high
sensitivity of these new TeV telescopes offer the possibility of
investigating the unidentified EGRET sources. For example, HESS is
an imaging telescope array system with an array of four imaging
atmospheric Cherenkov telescopes. Now it is fully operational, and
its expected sensitivity at 1 TeV is about $2\times 10^{-13}\ {\rm
photon\ cm^{-2}\ s^{-1}}$, then we also expect HESS could possibly
detect TeV photons from about 15 unidentified EGRET sources if
they are mature pulsars.

We thank the anonymous referee for his very useful comments and
Prof. P.K. MacKeown for his critical reading. This work is
supported by a RGC grant of the Hong Kong Government.

\begin{figure}
\psfig{figure=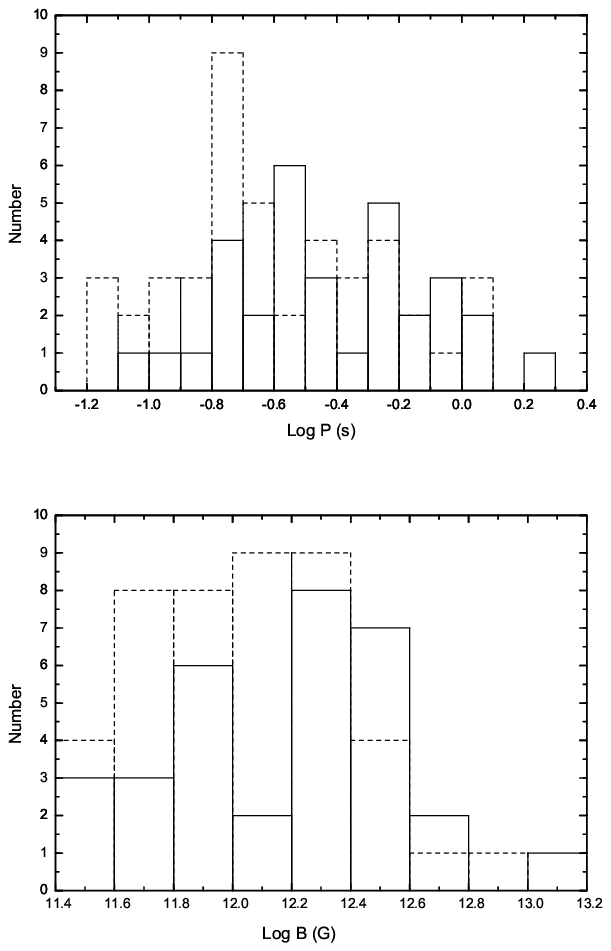,angle=0,width=10cm} \caption{The
distribution of the period (upper panel) and surface magnetic
field (bottom panel) of the simulated $\gamma$-ray pulsars which
could be detected by EGRET. The distributions of the pulsars in
the high latitude ($|b|>5^\circ$) (solid), and in the Galactic
disk ($|b|\leq 5^\circ$) (dashed) are shown.}
\end{figure}

\begin{figure}
\psfig{figure=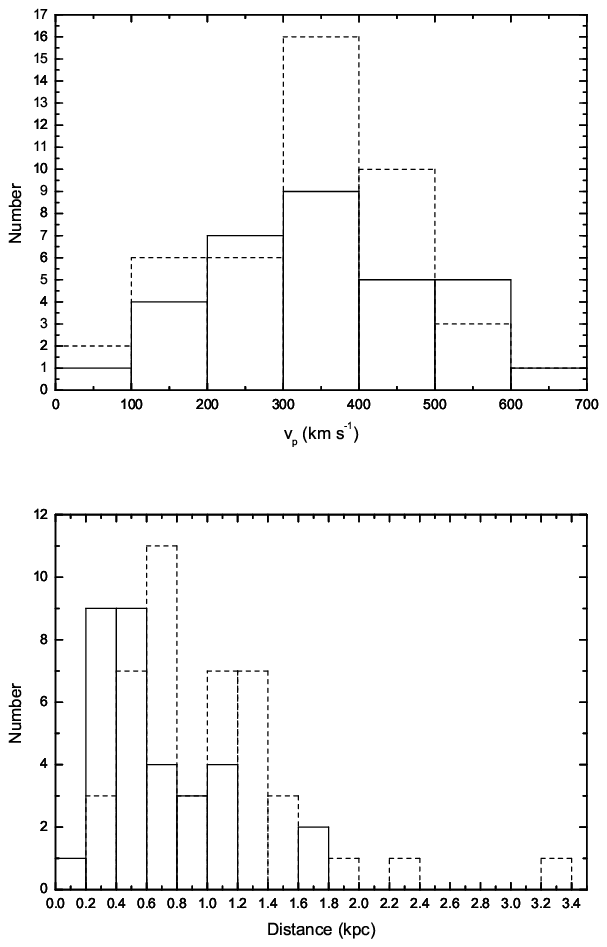,angle=0,width=10cm} \caption{The
distribution of the proper motion velocity (top) and the distance
(bottom) of the simulated $\gamma$-ray pulsars. The distributions
of the pulsars in the high latitude ($|b|>5^\circ$) (solid), and
in the Galactic disk ($|b|\leq 5^\circ$) (dashed) are shown.}
\end{figure}

\begin{figure}
\psfig{figure=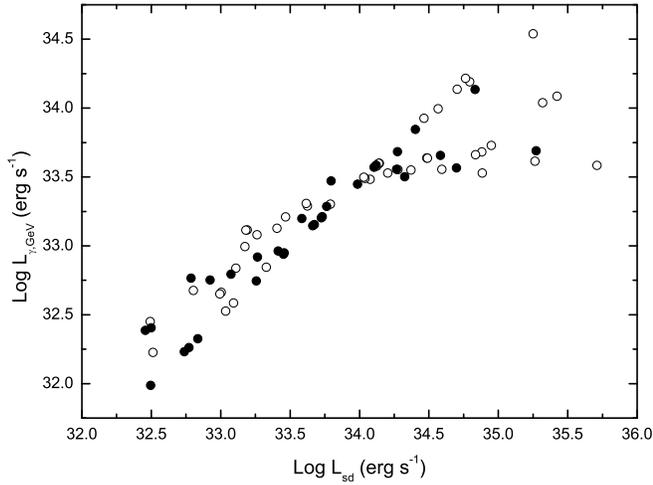,angle=0,width=10cm} \caption{The GeV
gamma-ray luminosity versus the spin down power for the simulated
$\gamma$-ray pulsars for the high latitude $|b|>5^\circ$ (solid)
and the Galactic latitude $|b|\leq 5^\circ$ (circle) sample.}
\end{figure}

\begin{figure}
\psfig{figure=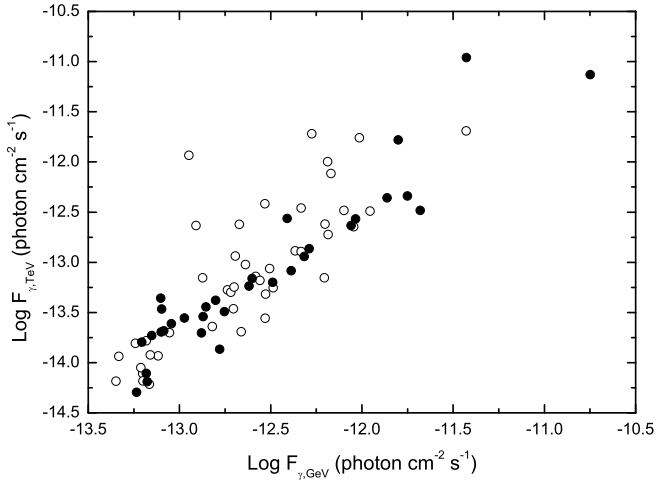,angle=0,width=10cm} \caption{The GeV
gamma-ray flux versus the TeV flux for the simulated $\gamma$-ray
pulsars for the high latitude $|b|>5^\circ$ (solid) and the
Galactic latitude $|b|\leq 5^\circ$ (circle) sample.}
\end{figure}

\begin{figure}
\psfig{figure=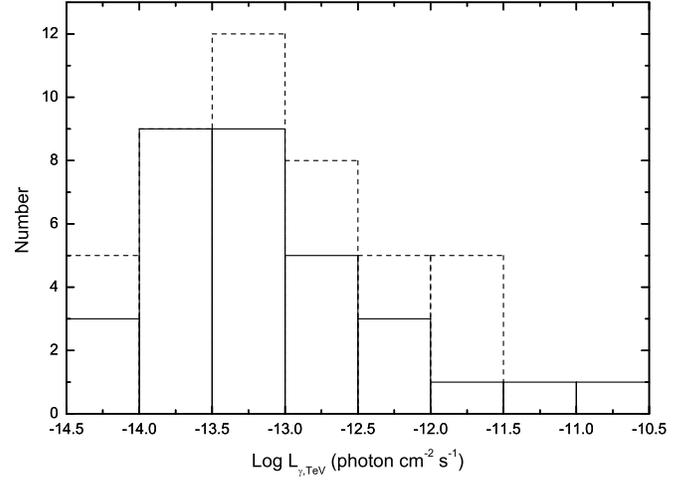,angle=0,width=10cm} \caption{The
distribution of the TeV flux from the wind nebulae of the
simulated $\gamma$-ray pulsars which could be the unidentified
EGRET sources. The distributions of the pulsars in the high
latitude ($|b|>5^\circ$) (solid), and in the Galactic disk
($|b|\leq 5^\circ$) (dashed) are shown.}
\end{figure}

\end{document}